\documentclass[aps,prl,showpacs,twocolumn,amsmath,amssymb]{revtex4} 
\usepackage{graphicx} 
\usepackage{Jonasmacros} 
\usepackage{bm} 
\usepackage[usenames,dvipsnames]{color}

\begin{document} 
\date{\today} 
\title{Atomistic spin dynamic method with both damping and moment of inertia effects included from first principles}
\author{S. Bhattacharjee} 
\author{L. Nordstr\"om}
\author{J. Fransson} 
\email{Jonas.Fransson@physics.uu.se} 
\affiliation{Department of Physics and Astronomy, Box 516, 75120, Uppsala University, Uppsala, Sweden}

%\author{A. V. Balatsky} 
%\email{avb@lanl.gov} 
%\affiliation{Theoretical Division, Los Alamos National Laboratory, Los Alamos, New Mexico 87545, USA} 
%\affiliation{Center for Integrated Nanotechnology, Los Alamos National Laboratory, Los Alamos, New Mexico 87545, USA} 
%\author{J. -X. Zhu} 
%\email{jxzhu@lanl.gov} 
%\affiliation{Theoretical Division, Los Alamos National Laboratory, Los Alamos, New Mexico 87545, USA} 
\begin{abstract}
We consider spin dynamics for implementation in an atomistic framework and we address the feasibility of capturing processes in the femtosecond regime by inclusion of moment of inertia. In the spirit of an {\it s-d} -like interaction between the magnetization and electron spin, we derive a generalized equation of motion for the magnetization dynamics in the semi-classical limit, which is non-local in both space and time. Using this result we retain a generalized Landau-Lifshitz-Gilbert equation, also including the moment of inertia, and demonstrate how the exchange interaction, damping, and moment of inertia, all can be calculated from first principles.
%The magnetization dynamics in the inertial regime is discussed from the point of view of dynamical equation consisting relaxation rate of faster environmental degrees of freedom coupled to {\it magnetic acceleration} or {\it retardation}. Using linear response formalism, we estimate the magnitude of the relaxation tensor associated with the faster environmental degrees of freedom. The diagonal components of this tensor describe the relaxation time for the remaining degrees of freedom beyond than that of the magnetization itself. We show that this relaxation time lies in femtosecond regime, and is related to the bandwidth of the material. Such a scaling obviously indicates a possible way of exploring magnetization dynamics at a time-scale much faster ({\it up to three orders of magnitudes}) than what is limited by Landau-Lifshitz-Gilbert formalism. 
\end{abstract} 
%73.63.Rt, 07.79.Cz, 72.25.Hg}

\maketitle

In recent years there has been a huge increase in the interest in fast magnetization processes on a femto-second scale, which has been initialized by important developments in experimental techniques \cite{beaurepaire1996,zhang2000,munzenberg2010,johnson2011,kirilyuk2010}, as well as potential technological applications \cite{akerman2005}. From a theoretical side, the otherwise trustworthy spin dynamical (SD) simulation method fails to treat this fast dynamics due to the short time and length scales involved. Attempts have been made to generalize the mesoscopic SD method to an atomistic SD, in which the dynamics of each individual atomic magnetic moment is treated \cite{nowak,skubic}. While this approach should in principle be well suited to simulate the fast dynamics observed in experiments, it has not yet reached full predictive power as it has inherited phenomenological parameters, e.g.~Gilbert damping, from the mesoscopic SD. The Gilbert damping parameter is well established in the latter regime but it is not totally clear how it should be transferred to the atomic regime. In addition, very recently it was pointed out that the moment of inertia, which typically is neglected, plays an important role for fast processes \cite{inertia}. In this Letter we derive the foundations for an atomistic SD where all the relevant parameters, such as the exchange coupling, Gilbert damping, and moment of inertia, can be calculated from first principles electronic structure methods.

Usually the spin dynamics is described by the phenomenological Landau-Lifshitz-Gilbert (LLG) 
equation \cite{ll,gil} which is composed of precessional and damping terms driving the dynamics to an equilibrium. By including the moment of inertia, we arrive at a generalized LLG equation
\begin{align} 
\dot\bfM=&
	\bfM\times(-\gamma\bfB+\hat{\bfG}\dot\bfM+\hat{\mathbf I}\ddot\bfM)
\label{eq-LLG}
\end{align}
where $\hat \bfG$ and $\hat{\mathbf I}$ are the Gilbert damping and the moment of inertia tensors, respectively. In this equation the effective field $\bfB$ includes both the external and internal fields, of which the latter includes the exchange coupling and anisotropy effects. Here, we will for convenience include the anisotropy arising from the classical dipole-dipole interaction responsible for the shape anisotropy as a part of the external field.  The damping term in the LLG equation usually consists of a single damping parameter, which essentially means that the time scales of the magnetization variables and the environmental variables are {\it well separated}. This separation naturally brings a limitation to the LLG equation concerning its time scale which is restricting it to the mesoscopic regime.

\begin{figure}[b]
\begin{center}
\includegraphics[width=0.99\columnwidth]{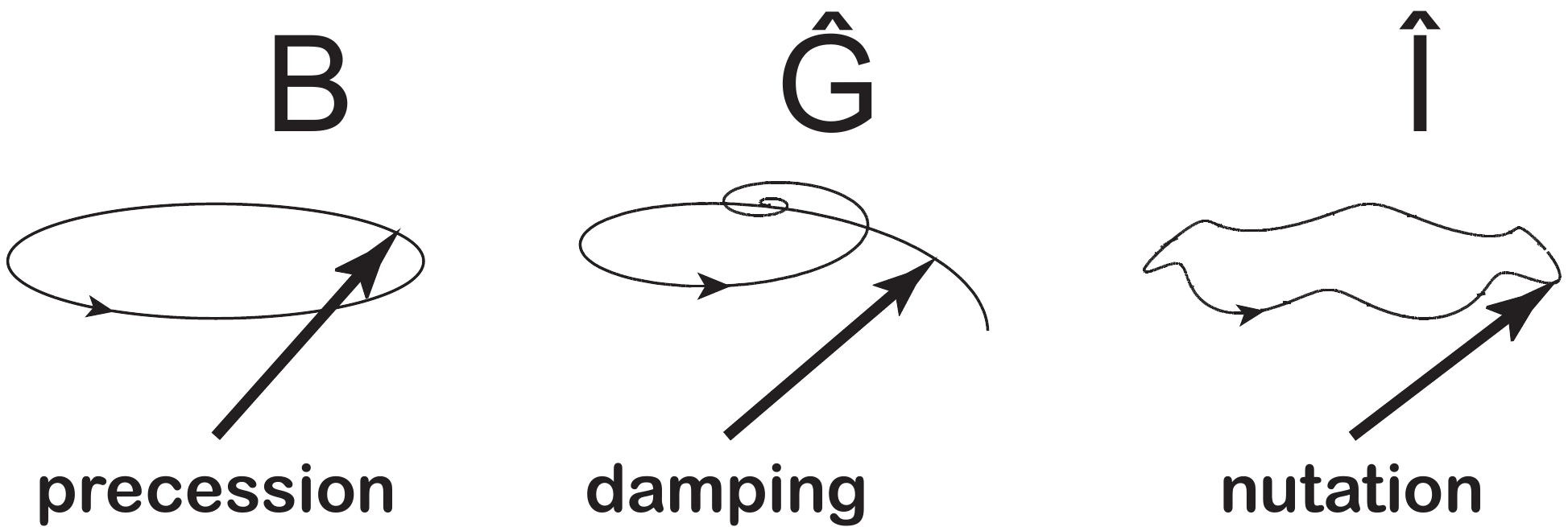}
\end{center}
\caption{The three contributions in Eq.~(\ref{eq-LLG}), the bare precession arising from the effective magnetic field, and the superimposed effects from the Gilbert damping and the moment of inertia, respectively.}
\label{fig-SD}
\end{figure}
The addition of a moment of inertia term to the LLG equation can be justifies as follows. A general process of a moment $\bfM$ under the influence of a field $\bfF$ is always endowed with inertial effects at higher frequencies \cite{sack}. The field  $\bfF$ and moment $\bfM$ can, for example, be stress and strain for mechanical relaxation, electric field and electric dipole moment in the case of dielectric relaxation, or magnetic field and magnetic moment in the case of magnetic relaxation. In this Letter we focus on the latter case | the origin of the moment of inertia in SD. The moment of inertia leads to nutations of the magnetic moments, see Fig. \ref{fig-SD}. Its wobbling variation of the azimuthal angle has a crucial role in fast SD, such as fast magnetization reversal processes.

In the case of dielectric relaxation the inertial effects are quite thoroughly mentioned in the 
literature \cite{Coffey,del2}, especially in the case of ferroelectric relaxors. Coffey {\it et al.} \cite{del2} have proposed inertia corrected Debye's theory of dielectric relaxation and showed that by including inertial effects, the unphysical high frequency divergence of the absorption co-efficient is removed.

Very recently Ciornei {\it et al} \cite{inertia} have extended the LLG equation to include the inertial effects through a {\it magnetic retardation} term in addition to precessional and damping terms. They considered a collection of uniformly magnetized particles and treated the total angular momentum $\bfL$ as faster variable. They obtained Eq.~(\ref{eq-LLG}) from a Fokker-Plank equation where the number density of magnetized particles were calculated by integrating a non-equilibrium distribution function over faster variables such that faster degrees of freedom appear as parameter in the calculation. 

The authors showed that at very short time scales the inertial effects become
important as the precessional motion of magnetic moment gets superimposed with nutation loops due to inertial effects. It is pointed out that the existence of inertia driven magnetization dynamics open up a pathway for ultrafast magnetic switching \cite{ultra} beyond the limitation \cite{brown} of the precessional switching.

In practice, to perform atomistic spin dynamics simulations the knowledge of $\hat \bfG$ and $\hat{\mathbf I}$ is necessary. There are recent proposals \cite{kelly,Gil} of how to calculate the Gilbert damping factor from first principles in terms of Kubo-Greenwood like formulas. Here, we show that similar techniques may by employed to calculate the moment of inertia tensor $\hat{\mathbf I}$. Finally, we present a microscopical justification of Eq.~(\ref{eq-LLG}), considering a collective magnetization density interacting locally with electrons constituting spin moments. Such a description would in principle be consistent with the study of magnetization dynamics where the exchange parameters are extracted from first-principles electronic structure calculations, e.g density functional theory (DFT) methods. We find that in an atomistic limit Eq.~(\ref{eq-LLG}) actually  has to be generalized slightly as both the damping and inertia tensors are naturally non-local in the same way as the exchange coupling included in the effective magnetic field $\bfB$. From our study it is clear that both the damping and the moment of inertia effects naturally  arise from the retarded exchange interaction.

We begin by considering the magnetic energy $E=\bfM\cdot\bfB$. Using that its time derivative is $\dot{E}=\bfM\cdot\dot\bfB+\dot\bfM\cdot \bfB$ along with Eq.~(\ref{eq-LLG}), we write
\begin{align} 
\dot{E}=&
	\bfM\cdot \dot\bfB
	+\frac{1}{\gamma}\dot\bfM\cdot
	\left(
		\hat{\bfG}\dot\bfM
		+\hat{\mathbf I}\ddot\bfM
	\right).
\label{eq-LLGder} 
\end{align}
Relating the rate of change of the total energy to the Hamiltonian $\Hamil$, through $\dot{E}=\av{d\Hamil/dt}$, and expanding the $\Hamil$ linearly around its static magnetization $\bfM_0$, with $\bfM(t)=\bfM_0+\mu(t)$, we can write $\Hamil\approx\Hamil_0+\mu(t)\cdot\nabla_\mu\Hamil_0$, where $\Hamil_0=\Hamil(\bfM_0)$. Then the rate of change of the total energy equals $\dot{E}=\dot{\mu}\cdot\av{\nabla_\mu\Hamil}$ to the first order. Following Ref.~\cite{brataas} and assuming sufficiently slow dynamics such that $\mu(t')=\mu(t)-\tau\dot\mu(t)+\tau^2\ddot\mu(t)/2$, $\tau=t-t'$, we can write the rate of change of the magnetic energy as
\begin{align} 
\dot{E}=& 
	\lim_{\omega\rightarrow0} 
	\dot\mu_i
	[
		\chi_{ij}(\omega)\mu_j 
		+i\partial_\omega\chi_{ij}(\omega)\dot\mu_j 
		-\partial^2_\omega\chi_{ij}(\omega)\ddot\mu_j/2
	].
\label{eq-pertder}
\end{align}
Here, $\chi_{ij}(\omega)=\int(-i)\theta(\tau)\av{\com{\partial_i\Hamil_0(t)}{\partial_j\Hamil_0(t')}}e^{i\omega\tau}dt'$, $\tau=t-t'$, is the (generalized) exchange interaction tensor out of which the damping and moments of inertia can be extracted. Summation over repeated indices ($i,j=x,y,z$) is assumed in Eq.~(\ref{eq-pertder}). Equating Eqs.~(\ref{eq-LLGder}) and (\ref{eq-pertder}) results in an internal contribution to the effective field about which the magnetization precesses $\bfB_\mathrm{int}=\mu\lim_{\omega\rightarrow0}\chi(\omega)$, the damping term $\hat{\bfG}=	\gamma\lim_{\omega\rightarrow0}i\partial_\omega\chi(\omega)$ as well as the moment of inertia $\hat{\mathbf I}=-\gamma\lim_{\omega\rightarrow0}\partial^2_\omega\chi(\omega)/2$.
 
For a simple order of magnitude estimate of the damping and inertial coefficients, $\hat{\bfG}$ and $\hat{\mathbf I}$, respectively, we may assume for a state  close to a ferromagnetic state that the spin resolved density of electron states $\rho_\sigma(\dote{})$ corresponding to the static magnetization configuration $\Hamil_0$ is slowly varying with energy. At low temperatures we, then, find
\begin{align}
\hat{\bfG}\sim&
	2\gamma\pi\,\mathrm{sp}
		\left[\av{\partial_iH_0}\rho
		\av{\partial_jH_0}
		\rho\right]_{\varepsilon=\varepsilon_\mathbf{F}}
\label{Gilb_approx},
\end{align}
in agreement with previous results \cite{brataas}. Here, sp denotes the trace over spin 1/2 space. By the same token, the moment of inertia is estimated as
\begin{align}
\hat{\mathbf I}\sim&
	-(\gamma/D)\,\mathrm{sp}
	\left[
		\av{\partial_iH_0}\rho
		\av{\partial_jH_0}
		\rho\right]_{\varepsilon=\varepsilon_\mathbf{F}},\label{est-MoI}
\end{align}
where $2D$ is the band width of density of electron states of the host material. Typically, for metallic systems the band width $2D\sim1$|$10$ eV, which sets the time-scale of the inertial contribution to the \emph{femto second} ($10^{-15}$ s) regime. It, therefore, defines magnetization dynamics on a time-scale that is one or more orders of magnitude shorter compared to e.g.~the precessional dynamics of the magnetic moment.

Next, we consider the physics leading to the LLG equation given in Eq.~(\ref{eq-LLG}). As there is hardly any microscopical derivation of the LLG equation in the literature, we include here, for completeness the arguments that leads to the equation for the spin-dynamics from a quantum field theory perspective.

In the atomic limit the spin degrees of freedom are deeply intertwined with the electronic degrees of freedom, and hence the main environmental coupling is the one to the electrons. In this study we are mainly concerned with a mean field description of the electron structure, as in the spirit of the DFT. Then a natural and quite general description of the magnetic interaction due to electron-electron interactions on the atomic site around $\bfr$ within the material is captured by the {\it s-d}-like model $\Hamil_\text{int}=-\int J(\bfr,\bfr')\bfM(\bfr,t)\cdot\bfs(\bfr',t)d\bfr d\bfr'$, where $J(\bfr,\bfr')$ represents the interaction between the magnetization density $\bfM$ and the electron spin $\bfs$.
%Here, $\bfs(\bfr,t)$ is defined in terms of the electron operators $\cdagger{}(\bfr,t)$ and $\cc{}(\bfr,t)$ according to $\bfs(\bfr,t)=(1/2)\sum_{\sigma\sigma'}\cdagger{}(\bfr,t)\bfsigma_{\sigma\sigma'}\cs{\sigma'}(\bfr,t)$.
From a DFT perspective the interaction parameter $J(\bfr,\bfr')$ is related to the effective spin dependent exchange-correlation functional $B_\mathrm{xc}[M(\bfr')](\bfr)$. For generality we assume a fully relativistic treatment of the electrons, i.e. including the spin-orbit coupling. In this interaction the dichotomy of the electrons is displayed, they both form the magnetic moments and provide the interaction among them.

Owing to the general non-equilibrium conditions in the system, we define the action variable
\begin{align} 
\calS=\oint_C\Hamil_\text{int}\,dt+\calS_\mathrm{Z}+\calS_\mathrm{WZWN}
\end{align}
on the Keldysh contour \cite{path1,path2,fransson2010}. Here, the action $\calS_\mathrm{Z}=-\gamma\oint_C\int\bfB_\mathrm{ext}(\bfr,t)\cdot\bfM(\bfr,t)dtd\bfr$ represents the Zeeman coupling to the external field $\bfB_\text{ext}(\bfr,t)$, whereas the Wess-Zumino-Witten-Novikov (WZWN) term $\calS_\mathrm{WZWN}=\int\oint_C\int_0^1\bfM(\bfr,t;\tau)\cdot[\partial_\tau\bfM(\bfr,t;\tau)\times\partial_t\bfM(\bfr,t;\tau)]d\tau dt|\bfM(\bfr)|^{-2}d\bfr$ describes the Berry phase accumulated by the magnetization.

In order to acquire an effective model for the magnetization density $\bfM(\bfr,t)$, we make a second order \cite{2ndorder} expansion of the partition function ${\cal Z}[\bfM(\bfr,t)]\equiv\tr T_Ce^{i\calS}$, and take the partial trace over the electronic degrees of freedom in the action variable. The effective action $\delta\calS_M$ for the magnetization dynamics arising from the magnetic interactions described in terms of $\Hamil_\text{int}$, can, thus, be written
\begin{align}
\delta\calS_M=&
	-
	\oint
		\int\bfM(\bfr,t)\cdot
			{\cal D}(\bfr,\bfr';t,t')
%\nonumber\\&
			\cdot\bfM(\bfr',t')
		d\bfr d\bfr'
	dtdt',
\end{align}
where ${\cal D}(\bfr,\bfr';t,t')=\int J(\bfr,\bfr_1)\eqgr{\bfs(\bfr_1,t)}{\bfs(\bfr_2,t')}\times J(\bfr_2,\bfr')d\bfr_1d\bfr_2$ is a dyadic which describes the electron mediated exchange interaction.

Conversion of the Keldysh contour integrations into real time integrals on the interval $(-\infty,\infty)$ results in
\begin{align}
\calS=&
	\int\bfM^{(\mathrm{fast})}(\bfr,t)\cdot[\bfM (\bfr,t)\times\dot\bfM (\bfr,t)]dt|\bfM(\bfr)|^{-2}d\bfr
\nonumber\\&
	+\int
		\bfM^{(\mathrm{fast})}(\bfr,t)\cdot{\cal D}^r(\bfr,\bfr';t,t')\cdot\bfM (\bfr',t')
	d\bfr  d\bfr'dtdt'
\nonumber\\&
	-\gamma \int\bfB_\text{ext}(\bfr,t)\cdot\bfM^{(\mathrm{fast})}(\bfr,t)dtd\bfr
	,
\label{gen-S}
\end{align}
with $\bfM^{(\mathrm{fast})}(\bfr,t)=\bfM_u(\bfr,t)-\bfM_l(\bfr,t)$ and $\bfM (\bfr,t)=[\bfM_u(\bfr,t)+\bfM_l(\bfr,t)]/2$ which define fast and slow variables, respectively. Here, $\bfM_{u(l)}$ is the magnetization density defined on the upper (lower) branch of the Keldysh contour. Notice that upon conversion into the real time domain, the contour ordered propagator ${\cal D}$ is replaced by its retarded counterpart ${\cal D}^r$.

We obtain the equation of motion for the (slow) magnetization variable $\bfM(\bfr,t)$ in the classical limit by minimizing the action with respect to $\bfM^{(\mathrm{fast})}(\bfr,t)$, cross multiplying by $\bfM(\bfr,t)$ under the assumption that the total moment is kept constant. We, thus, find
\begin{align}
\dot\bfM(\bfr,t)=&
	\bfM(\bfr,t)\times
	\biggl(
		-\gamma\bfB_\mathrm{ext}(\bfr,t)
\nonumber\\&
		+
		\int
			{\cal D}^r(\bfr,\bfr';t,t')\cdot\bfM(\bfr',t')dt'd\bfr'
	\biggr).
\label{gen-DM}
\end{align}
%\begin{align}
%-\dot\bfM&
%	+\frac{1}{|\bfM|^2}\biggl(\bfM[\bfM\cdot\dot{\bfM}]+i\bfM\times\dot{\bfM}\biggr)
%%\nonumber\\&
%	+\gamma\bfB_\text{ext}\times\bfM
%\nonumber\\&
%	+\bfM\times\int{\cal D}^r(\bfr,\bfr';t,t')\cdot\bfM(\bfr',t')dt' d\bfr'
%	=0,
%\label{eq-QLLG}
%\end{align}
%where we have suppressed the dependence on $(\bfr,t)$ on some quantities.
%Notice that this result is still entirely quantum mechanical \comment{not true for this equation since the assumptions of constant $\bfM$ and commuting magnetization densities have been applied} and would, hence, be applicable for time-scales much shorter than what is typically considered in spin dynamical systems. In this expression the time derivative $\dot\bfM$ is the only non-vanishing term from the triple cross product $\bfM\times(\bfM\times\dot\bfM)=\bfM(\bfM\cdot\dot\bfM)-|\bfM|^2\dot\bfM+i\bfM\times\dot\bfM$ stemming from the first term in the action of Eq.~(\ref{gen-S}). As for the other two terms the last contribution, $i\bfM\times\dot\bfM$, is solely due to the non-commutativity of quantum spins and vanish in the classical limit, while the first term is zero for purely transversal fluctuations, e.g.~when the total moment $\bfM$ is kept constant which leads to that $\dot\bfM\cdot\bfM=\bfM\cdot\dot\bfM=(1/2)\dt|\bfM|^2=0$.
%The equation for the magnetization dynamics can, hence, be cast in the form
%\begin{align}
%\dot\bfM(\bfr,t)=&
%	\bfM(\bfr,t)\times
%	\biggl(
%		-\gamma\bfB_\mathrm{ext}(\bfr,t)
%\nonumber\\&
%		+\int
%			{\cal D}^r(\bfr,\bfr';t,t')\cdot\bfM(\bfr',t')dt'd\bfr'
%	\biggr).\label{gen-DM}
%\end{align}
Eq.~(\ref{gen-DM}) provides a generalized description of the semi-classical magnetization dynamics compared to the LLG Eq.~(\ref{eq-LLG}) in the sense that it is non-local in both time and space. The dynamics of the magnetization at some point $\bfr$ depends not only on the magnetization locally at $\bfr$, but also in a non-trivial way on the surrounding magnetization. The coupling of the magnetization at different positions in space is mediated via the electrons in the host material. Moreover, the magnetization dynamics is, in general, a truly non-adiabatic process in which the information about the past is crucial.

However, in order to make connection to the magnetization dynamics as described by e.g.~the LLG equation as well as Eq.~(\ref{eq-LLG}) above, we make the following consideration. Assuming that the magnetization dynamics is slow compared to the electronic processes involved in the time-non-local field ${\cal D}(\bfr,\bfr';t,t')$, we expand the magnetization in time according to $\bfM(\bfr',t')\approx\bfM(\bfr',t)-\tau\dot\bfM(\bfr',t)+\tau^2\ddot\bfM(\bfr',t)/2$. Then for the integrand in Eq.~(\ref{gen-DM}), we get 
\begin{align} 
\lefteqn{
	{\cal D}^r(\bfr,\bfr';t,t')\cdot\bfM(\bfr',t')=}
\nonumber\\&
	{\cal D}^r(\bfr,\bfr';t,t')\cdot[\bfM(\bfr',t)
	-\tau\dot\bfM(\bfr',t)+\frac{\tau^{2}}{2}\ddot\bfM(\bfr',t)].
\label{eq-Integrand}
\end{align} 
Here, we observe that as the exchange coupling for the magnetization is non-local and mediated through ${\cal D}$, this is also true for the damping (second term) and the inertia (third term).

In order to obtain an equation of the exact same  form as LLG in Eq.~(\ref{eq-LLG}) we further have to assume that the magnetization is close to a uniform ferromagnetic state, then we can justify the approximations $\dot\bfM(\bfr',t)\approx\dot\bfM(\bfr,t)$ and $\ddot\bfM(\bfr',t)\approx\ddot\bfM(\bfr,t)$. When $\bfB_\mathrm{int}=-\int{\cal D}(\bfr,\bfr';t,t')\cdot\bfM(\bfr',t)d\bfr'dt'/\gamma$ is included in the total effective magnetic field $\bfB$, the tensors of Eq.~(\ref{eq-LLG}) $\hat\bfG$ and $\hat{\mathbf I}$ can be identified with $-\int\tau{\cal D}(\bfr,\bfr';t,t')d\bfr'dt'$ and $\int\tau^2{\cal D}(\bfr,\bfr';t,t')d\bfr'dt'/2$, respectively. From a first principles model of the host materials we have, thus, derived the equation for the magnetization dynamics discussed in Ref.~\onlinecite{inertia}, where it was considered from purely classical grounds. However it is clear that for a treatment of atomistic SD that allows for all kinds of magnetic orders, not only ferromagnetic, Eq.~(\ref{eq-LLG}) is not sufficient and  the more general LLG equation of Eq.~(\ref{gen-DM}) together with Eq.~(\ref{eq-Integrand}) has to be used.

We finally describe how the parameters of Eq.~(\ref{eq-LLG}) can be calculated from a first principles point of view. Within the conditions defined by the DFT system, the interaction tensor ${\cal D}^r$ is time local which allows us to write $\lim_{\dote{}\rightarrow0}i\partial_{\dote{}}{\cal D}^r(\bfr,\bfr';\dote{})=\int\tau{\cal D}^r(\bfr,\bfr';t,t')dt'$ and $\lim_{\dote{}\rightarrow0}\partial^2_{\dote{}}{\cal D}^r(\bfr,\bfr';\dote{})=-\int\tau^2{\cal D}^r(\bfr,\bfr';t,t')dt'$, where
\begin{align}
{\cal D}^r&(\bfr,\bfr';\dote{})=
	4\,\sp
		\int
			J_{\bfr\bfrho}
			J_{\bfrho'\bfr'}
			\frac{f(\omega)-f(\omega')}{\dote{}-\omega+\omega'+i\delta}
%			\bfsigma
%			\Bigl[
%				\bfG^<(\bfr_2,\bfr_1;\omega)\bfsigma\bfG^>(\bfr_1,\bfr_2;\omega-\dote{})
%\nonumber\\&
%				-\bfG^>(\bfr_2,\bfr_1;\omega)\bfsigma\bfG^<(\bfr_1,\bfr_2;\omega-\dote{})
%			\Bigr]
\nonumber\\&\times
			\bfsigma\im\bfG^r_{\bfrho'\bfrho}(\omega)\bfsigma\im\bfG^r_{\bfrho\bfrho'}(\omega')
%\nonumber\\&\times
		\frac{d\omega}{2\pi}
		\frac{d\omega'}{2\pi}
		d\bfrho d\bfrho'.
\label{eq-D}
\end{align}
Here, $J_{\bfr\bfr'}\equiv J(\bfr,\bfr')$ whereas $\bfG^r_{\bfr\bfr'}(\omega)\equiv\bfG^r(\bfr,\bfr';\omega)$ is the retarded GF, represented as a $2\times2$-matrix in spin-spaces. We notice that the above result presents a general expression for frequency dependent exchange interaction. Using Kramers-Kr\"onig's relations in the limit $\dote{}\rightarrow0$, it is easy to see that Eq.~(\ref{eq-D}) leads to
\begin{align}
{\cal D}^r(\bfr,\bfr';0)=&
	-\frac{1}{\pi}
	\,\sp\,\im
		\int
			J_{\bfr\bfrho}J_{\bfrho'\bfr'}
			f(\omega)
\nonumber\\&\times
			\bfsigma\bfG^r_{\bfrho'\bfrho}(\omega)\bfsigma\bfG^r_{\bfrho\bfrho'}(\omega)
		d\omega d\bfrho d\bfrho',
\label{eq-exch}
\end{align}
in agreement with e.g.~Ref.~\cite{antropov1995}.  We can make connection with previous results, e.g.~Refs.~\onlinecite{antropov1997,katsnelson2004}, and observe that Eq.~(\ref{eq-D}) contains the isotropic Heisenberg, anisotropic Ising, and Dzyaloshinsky-Moriya exchange interactions between the magnetization densities at different points in space \cite{fransson2010}, as well as the onsite contribution to the magnetic anisotropy.

Using the result in Eq.~(\ref{eq-D}), we find that the damping tensor is naturally non-local and can be reduced to
\begin{align}
\hat{\bfG}(\bfr,\bfr')=&
	\frac{1}{\pi}\,\sp\int
			J_{\bfr\bfrho}J_{\bfrho'\bfr'}
			f'(\omega)
\nonumber\\&\times
			\bfsigma\im\bfG^r_{\bfrho'\bfrho}(\omega)
			\bfsigma\im\bfG^r_{\bfrho\bfrho'}(\omega)
		d\omega d\bfrho d\bfrho'\,,
\label{eq-damping}
\end{align}
which besides the non-locality is in good accordance with the results in Refs.~\cite{antropov1997,kelly}, and is closely connected to the so-called torque-torque correlation model \cite{Kamb}. With inclusion of the the spin-orbit coupling in $\bfG^{r}$, it has been demonstrated that Eq.~(\ref{eq-damping}) leads to a local Gilbert damping of the correct order of magnitude for the case of ferromagnetic permalloys \cite{kelly}.

Another application of Kramers-Kr\"onig's relations leads, after some algebra, to the moment of inertia tensor
\begin{align}
\hat{\mathbf I}(\bfr,\bfr')=&
	\,\sp\int
		J_{\bfr\bfrho}J_{\bfrho'\bfr'}
		f(\omega)
		\bfsigma
		[
			\im\bfG^r_{\bfrho'\bfrho}(\omega)\bfsigma\partial_\omega^2\re\bfG^r_{\bfrho\bfrho'}(\omega)
\nonumber\\&
			+\im\bfG^r_{\bfrho\bfrho'}(\omega)\bfsigma\partial_\omega^2\re\bfG^r_{\bfrho'\bfrho}(\omega)
		]
		\frac{d\omega}{2\pi}
		d\bfrho d\bfrho',
\end{align}
where we notice that the moment of inertia is not simply a Fermi surface effect but depends on the electronic structure as a whole of the host material. Although the structure of this expression is in line with the exchange coupling in Eq.~(\ref{eq-exch}) and the damping of Eq.~(\ref{eq-damping}), it is a little more cumbersome to compute due the presence of the derivatives of the Green's functions. Note that it is not possible to get completely rid of the derivatives through partial integration. These derivatives also make the moment of inertia very sensitive to details of the electronic structure, which has a few implications. Firstly the moment of inertia can take large values for narrow band magnetic materials, such as strongly correlated electron systems, where these derivatives are substantial. For such systems the action of moment of inertia can be important for longer time scales too, as indicated by Eq.~(\ref{est-MoI}). Secondly, the moment of inertia may be strongly dependent on the reference magnetic ordering for which it is calculated.  It is well known that already the exchange tensor parameters may depend on the magnetic order.  It is the task of future studies to determine how transferable the moment of inertia tensor as well as damping tensor are in-between different magnetic ordering.

In conclusion, we have derived a method for atomistic spin dynamics which would be applicable for ultrafast (femtosecond) processes. Using a general \emph{s-d}-like interaction between the magnetization density and electron spin, we show that magnetization couples to the surrounding in a non-adiabatic fashion, something which will allow for studies of general magnetic orders on an atomistic level, not only ferromagnetic. By showing that our method capture previous formulas for the exchange interaction and damping tensor parameter, we also derive a formula for calculating the moment of inertia from first principles. In addition our results point out that all parameters are non-local as they enter naturally as bilinear sums in the same fashion as the well established exchange coupling. Our results are straight-forward to implement in existing atomistic SD codes, so we look on with anticipation to the first applications of the presented theory which would be fully parameter-free and hence can take a large step towards simulations with predictive capacity.
 
Support from the Swedish Research Council is acknowledged. We are grateful for fruitful and encouraging discussions with A. Bergman, L. Bergqvist, O. Eriksson, C. Etz, B. Sanyal, and A. Taroni. J.F. also acknowledges discussions with J. -X. Zhu.


\begin{thebibliography}{99}
\bibitem{2ndorder} If higher order spin interaction terms than bilinear are needed it is straight forward to include higher order in this expansion.
\end{thebibliography}
\end{document}